\documentclass[aps,prb,twocolumn,groupedaddress]{revtex4}
\usepackage{graphicx}% Include figure files
\usepackage{dcolumn}% Align table columns on decimal point

\begin{document}

%\preprint{}

\title{NMR and M\"ossbauer study of spin dynamics and electronic 
structure of 
Fe$_{2+x}$V$_{1-x}$Al and Fe$_{2}$VGa}


\author{C. S. Lue}
\author{Yang Li}
\author{Joseph H. Ross, Jr.}
%\thanks{}
\affiliation{Department of Physics, Texas A\&M University, College 
Station, TX 77843-4242}
\author{George M. Irwin}
%\thanks{}
\affiliation{Department of Chemistry \& Physics, Lamar University, 
Beaumont, TX 77710}


\date{\today}

\begin{abstract}
In order to assess the magnetic ordering process in Fe$_{2}$VAl and 
the related material Fe$_{2}$VGa, we have carried out nuclear 
magnetic resonance (NMR) and M\"ossbauer studies.  
$^{27}$Al NMR relaxation measurements covered the temperature range 
4 -- 500 K in Fe$_{2+x}$V$_{1-x}$Al samples.  We found a peak in the
NMR spin-lattice relaxation rate, $^{27}$({\it T}$_{1}^{-1}$), 
corresponding to the magnetic 
transitions in each of these samples.  These peaks appear at 125 K, 
17 K, and 165 K for 
{\it x } = 0.10, 0, and - 0.05 respectively, and we connect these
features with critical slowing down of the localized
antisite
defects. M\"ossbauer measurements for Fe$_{2}$VAl and 
Fe$_{2}$VGa showed lines with no hyperfine splitting, and isomer
shifts nearly identical to those of the corresponding sites in 
Fe$_{3}$Al and 
Fe$_{3}$Ga, respectively.  We show that a model in which local band
filling leads to magnetic regions in the samples, in addition to the
localized antisite defects, can account for the observed magnetic
ordering behavior.
\end{abstract}

\pacs{75.20.En, 76.60.Jx, 76.80.+y, 71.20.Lp}
%\keywords{}

\maketitle

%\section{}
%\subsection{}
%\subsubsection{}
\section{Introduction}

In recent years, much attention has been focused on inhomogeneous
magnetic materials and on their implications in both technological 
applications and 
fundamental physics. Granular ferromagnets have been widely 
investigated since the discovery of giant magnetoresistance (GMR) in 
these 
alloys.\cite{berkowitz92,xiao92} GMR in granular alloys is associated 
with 
the spin-dependent scattering of conduction 
electrons by disordered magnetic clusters, while in ordered
compounds GMR may be associated with scattering from thermal 
disorder near T$_{c}$.\cite{coey95}  The Heusler alloys 
Fe$_{2+x}$V$_{1-x}$Al
and Fe$_{2+x}$V$_{1-x}$Ga are ordered intermetallics which 
exhibit GMR near T$_{c}$,\cite{endo98,matsuda00} although 
in addition to thermal disorder, localized magnetic clusters
play a prominent role in these materials, even for the nominally
nonmagnetic composition $x$ = 0.\cite{lue99,lue01} There has been 
considerable interest in understanding the role of these clusters
for the magnetic transitions,
and the unusual electronic behavior associated with these
materials.\cite{nishino97,kato00,maksimov01}

Both Fe$_{2}$VAl and Fe$_{2}$VGa
have been shown to be semimetallic and intrinsically non-magnetic,
via electronic
structure calculations\cite{guo98,singh98,weht98,weinert98,bansil99}
NMR measurements,\cite{lue98,lue98rem,lue01} and optical 
measurements.\cite{okamura00,feng01} These materials adopt 
the Heusler structure (L2$_{1}$, or cF16 \#225, AlCu$_{2}$Mn type), 
having a 
BCC-based lattice in which each 
Fe (8$c$ position) has four V and four Al or Ga near-neighbors.  
Complete substitution by Fe on the
V sites (4$b$ positions) produces the Fe$_{3}$Al structure,
DO$_{3}$.  Fe$_{3}$Al is a metallic ferromagnet in which both Fe 
sites carry a 
magnetic moment.\cite{stearns68} In Fe$_{2+x}$V$_{1-x}$Al T$_{c}$ 
goes monotonically to zero as $x$ goes to zero,\cite{kato00} although 
Fe$_{2}$VAl exhibits a sample-dependent low-temperature transition
which may be residual ferromagnetism\cite{maksimov01,matsushita99} or
a superparamagnetic freezing temperature.\cite{lue01,feng01} 
Furthermore
in Fe$_{2}$VAl and Fe$_{2}$VGa localized magnetic defects have 
been observed and associated with Fe 
antisites,\cite{lue99,singh98,bansil99} while there is also evidence 
for
larger magnetic clusters in the materials.\cite{lue01,feng01}  No
magnetic splitting was observed in M\"ossbauer studies near the
Fe$_{2}$VAl composition,\cite{popiel89,shobaki00,maksimov01} while
the dominance of defects and clusters makes a percolative behavior
seem likely.

In order to better understand the local magnetic properties and their
changes with 
composition, we have undertaken NMR and M\"ossbauer shift
studies, providing information about on-site
electron densities and the local magnetic configurations 
of Fe$_{2}$VAl, Fe$_{2}$VGa, and the mixed composition 
Fe$_{2+x}$V$_{1-x}$Al.  In previous NMR studies
reported by two of us,\cite{lue98,lue00} the
lineshapes were shown to be sensitive probes of the 
magnetic defects in Fe$_{2}$VAl and Fe$_{2}$VGa, 
while $^{51}$V Korringa relaxation was used to characterize the
electronic structure in the region of the semimetallic gap.
Here we report measurements of non-Korringa $^{27}$Al
relaxation behavior near T$_{c}$ in
Fe$_{2+x}$V$_{1-x}$Al, for compositions bracketing
the $x=0$ phase.  We show that relaxation peaks near T$_{c}$,
together with the lineshapes for these compositions, give a 
consistent picture of the behavior of antisite defects in this
material.  The relaxation peaks are due to the critical slowing
down of these magnetic defects, and we show that the magnetic
defect density tracks the Fe content for the Fe-rich composition.
Furthermore, in
$^{57}$Fe M\"ossbauer results we identify the isomer shifts for
8$c$ sites in Fe$_{2}$VAl and Fe$_{2}$VGa to differ very little
from those of the corresponding Fe sites in 
Fe$_{3}$Al and metastable Fe$_{3}$Ga. Thus while the 4$b$-site
Fe content changes, the local electronic structure is relatively
little affected by the composition change.

\section{Experiment}

Samples used for all measurements were a
portion of the same ingots used in previous NMR\cite{lue98,lue00,lue01} and 
specific heat\cite{lue99} measurements. These are polycrystalline samples
prepared by arc melting. Ingots were annealed
in vacuum at 800 or 1000 $^\circ$C, 
then 400 $^\circ$C, followed by furnace cooling. 
X-ray powder analysis on all 
samples showed the expected L2$_{1}$ structure, and microprobe
analysis confirmed these to be single-phase.  Larger lattice 
constants were found for the non-stoichiometric 
Fe$_{2+x}$V$_{1-x}$Al compositions, similar to those reported in the 
literature.\cite{nishino97,kato00,zarek00} 

NMR experiments were performed at fixed field using a 9-T home-built 
pulse NMR spectrometer described elsewhere.\cite{lue98} 
$^{27}$Al NMR spectra for all samples were detected near 99 MHz. 
Ingots were powdered to an approximate 100 - 200 
$\mu$m particle size. 
These powders mixed with granular quartz were 
placed in thin walled plastic vials for 4 to 300 K measurements, 
and Teflon tubes for high temperature purposes. Both 
sample holders show no observable $^{27}$Al NMR signals. 

$^{57}$Fe M\"ossbauer measurements were obtained using a $^{57}$Co
source in Pd matrix, at ambient temperature, driven in the
triangle mode.  Results were least-squares fitted 
using Voigt lineshapes, with shifts referenced to ${\alpha}$-Fe.
The instrumental broadening was small, as
estimated from a Sodium Nitroprusside calibration sample; the reported
linewidths were corrected for this term.

\section{Results}

\begin{figure}
\includegraphics{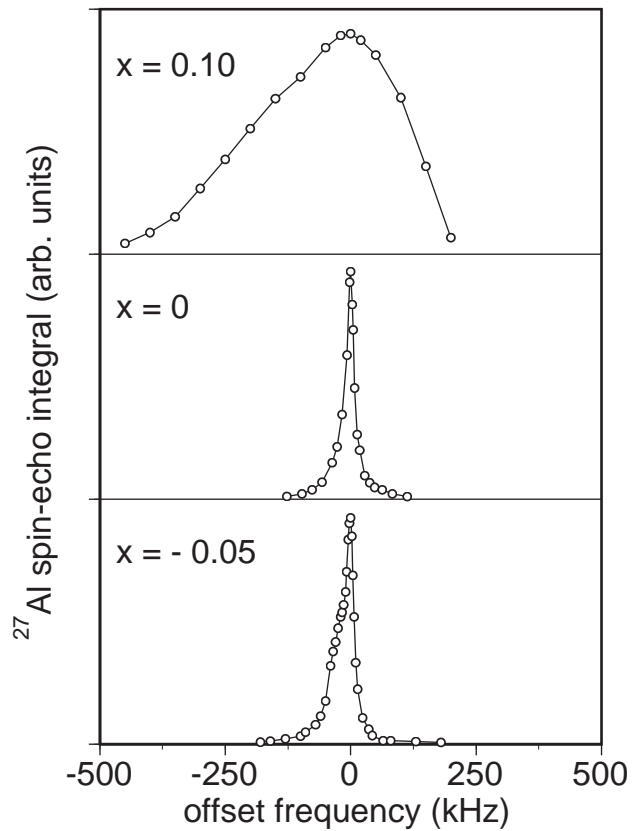}
\caption{\label{fig:fig1}Room temperature $^{27}$Al NMR powder 
patterns for 
Fe$_{2+x}$V$_{1-x}$Al with x = 0.10, 0, and - 0.05 respectively.}
\end{figure}

$^{27}$Al room-temperature NMR powder patterns for the 
Fe$_{2+x}$V$_{1-x}$Al compounds are shown in 
Fig.~\ref{fig:fig1}, 
measured by spin-echo integration versus frequency. We found 
relatively 
narrow NMR linewidths for Fe$_{2}$VAl and Fe$_{1.95}$V$_{1.05}$Al, 
although the latter exhibits an anisotropic line, 
presumably due to 
different neighbor configurations in the mixed alloy. The 
Fe$_{2.1}$V$_{0.9}$Al compound, however, shows a significantly 
broadened 
$^{27}$Al spectrum, which can be attributed to strong local 
magnetism, corresponding to the bulk magnetism observed 
for $x >$ 0.\cite{nishino97,kato00}  

The 4$a$ Al sites in Fe$_{2}$VAl have cubic site symmetry, and hence
no quadrupole broadening for $^{27}$Al NMR in the ideal
case.  However, from pulse-angle studies we determined that the 
satellite transitions are
broadened and made unobservable by random defects, leaving a
central line only (1/2 ${\leftrightarrow}$ -1/2 transition).  
This is typical behavior for dilute 
alloys,\cite{kanert71} even for the small defect
densities found in stoichiometric Fe$_{2}$VAl. 
In a previous NMR study\cite{lue01} of Fe$_{2}$VAl
(same ingot as the present $x$ = 0 sample),
$^{27}$Al and $^{51}$V linewidths were found to be identical,
and to have a Curie-law temperature dependence with a limiting 
high-temperature intrinsic width of 3 kHz.  The 
temperature-dependence is characteristic of broadening by
local moments, and was shown to be consistent with the presence
of magnetic antisite defects.\cite{lue01}
Clearly for the $x \neq 0$ cases the larger
linewidths (Fig.~\ref{fig:fig1}) must be caused by inhomogeneous
broadening due to the larger defect density in these alloys.  
From limited temperature-dependence studies we observed the 
non-stoichiometric samples to have 
linewidths increasing with decreasing 
temperature, as would be expected. 
We measured the spin-echo $T_{2}$ and observed a weak
temperature increase at low temperatures, to a value of 
210 $\mu$s at 4 K for $x$=0, consistent with the freezing-in
of local fields as the system goes through its spin freezing 
at low temperatures.\cite{lue01}  Maksimov et al.\cite{maksimov01}
have similarly seen a suppression of dynamic relaxation processes
at the spin freezing point, in a $\mu$SR study.

\begin{figure}
\includegraphics{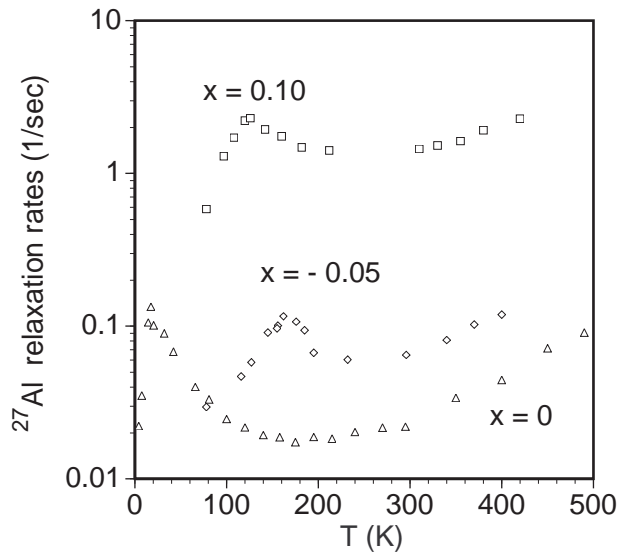}
\caption{\label{fig:fig2}Temperature dependence of $^{27}$Al 
relaxation rates in 
Fe$_{2+x}$V$_{1-x}$Al, indicated by triangles, circles, and diamonds 
for 
{\it x} = 0.10, 0, and - 0.05 respectively.}
\end{figure}

The antisite-defect dynamics are manifested much more clearly in
the $^{27}$Al $T_{1}$, rather than the $T_{2}$, 
and this provides
a powerful tool to investigate the spin dynamics. The $^{51}$V 
{\it T}$_{1}$ exhibits Korringa behavior at low temperatures,
indicating that conduction electrons dominate the relaxation of
this nucleus.  However, the $^{27}$Al $T_{1}$ was previously noted
to exhibit non-Korringa relaxation 
in Fe$_{2}$VAl.\cite{lue98,lue00} 
Here we report results of a detailed study showing a 
clear $T_{1}^{-1}$ peak for Fe$_{2}$VAl, and similar behavior in
the $x \neq 0$ materials. Results are shown in 
Fig.~\ref{fig:fig2} by squares, 
triangles, and diamonds, for 
{\it x} = 0.10, 0, and $-$0.05 respectively.
Rates were measured by inversion 
recovery, using the integral of the spin 
echo fast Fourier transform, irradiating the central portion 
of the $^{27}$Al line. For the recovery of the central 
transition, the {\it T$_{1}$}'s were extracted by fitting to 
multi-exponential curves\cite{narath67} appropriate for 
magnetic relaxation of the $I$ = 5/2 $^{27}$Al central transition. 

$T_{1}^{-1}$ maxima indicate the critical slowing down
of the local field fluctuations at the nuclear 
sites.\cite{moriyabook,narath72} We find that 
$^{27}$($T_{1}^{-1}$) passes through a maximum at 125 K, 17 
K, and 165 K for Fe$_{2.1}$V$_{0.9}$Al, Fe$_{2}$VAl and 
Fe$_{1.95}$V$_{1.05}$Al respectively. On further cooling, 
$^{27}$($T_{1}^{-1}$) drops rapidly in all samples. These peaks
correspond to susceptibility peaks observed in the
same samples.\cite{lue01} For Fe$_{2.1}$V$_{0.9}$Al this
peak corresponds to a ferromagnetic ordering temperature, while
for Fe$_{2}$VAl the transition may be a superparamagnetic 
freezing temperature.\cite{lue01} It is not entirely clear
why no peak was observed for $^{51}$($T_{1}^{-1}$), although
for the latter, the presence of $d$ orbitals significantly enhances the
Korringa relaxation process, while a small RKKY term might also
contribute and be important for $^{51}$V. 

\begin{figure}
\includegraphics{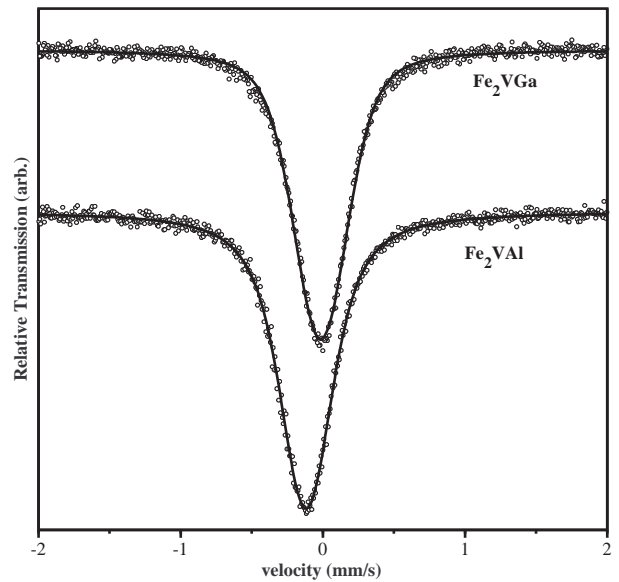}
\caption{\label{fig:fig3}Room-temperature M\"ossbauer spectra 
for Fe$_{2}$VAl and 
Fe$_{2}$VGa.  Least-squares fits are superposed on the data.  
Velocities are relative to the source.}
\end{figure}

Room-temperature M\"ossbauer measurements for both Fe$_{2}$VAl 
and Fe$_{2}$VGa yielded 
single lines, as shown in Fig.~\ref{fig:fig3}. The data are shown
in the figure, along with the least-squares fitting curves.  In both
cases, a good fit was obtained using a single Voigt-broadened line.
The two linewidths thus determined are identical, and slightly larger
than the natural width for Fe. Parameters
from these fits are given in Table~\ref{tab:table1}. The widths given
in the table are excess values over the natural width for Fe.  
These small widths indicate a relative lack of inhomogeneous 
broadening, showing these samples to be well-ordered in the 
Heusler structure, with very similar Fe site occupancies.
Also shown in
the table are isomer shifts for Fe$_{3}$Al and Fe$_{3}$Ga, from
Lin {\it et al.}\cite{lin81} and 
Kawamiya {\it et al.}\cite{kawamiya72}, respectively.  There 
have been a number of M\"ossbauer studies of the stable compound
Fe$_{3}$Al,\cite{stearns68,lin81,raju79,fultz94} while for 
Fe$_{3}$Ga the DO$_{3}$ structure is metastable, and quenched
samples were used to obtain the referenced values.  Finally, for
comparison, values for the vanadium Knight shift and the cubic
lattice constant of Fe$_{2}$VAl and Fe$_{2}$VGa are shown in the 
table.  These values are quite
close, indicating the similarity of these two materials.

\begin{table}
\caption{\label{tab:table1}Measured M\"ossbauer isomer
shifts and widths, along with NMR Knight shifts and lattice
parameters. Room-temperature isomer shifts ($\delta$)
and line widths ($\sigma$) are listed.  Isomer shifts are quoted
relative to an ${\alpha}$-Fe reference. Vanadium Knight shifts ($K_{V}$)
are low-temperature values.  Wyckoff positions labels
are given for the DO$_{3}$ lattice (equivalent to the L2$_{1}$
labels).}
\begin{ruledtabular}
\begin{tabular}{ccclcc}
Material&\mbox{a (nm)}&site&${\delta}$ (mm/s)&${\sigma}$ 
(mm/s)&K$_{V}$ (\%)\\ 
\hline
Fe$_{2}$VAl&0.576\footnotemark[2]&8$c$&0.058(5)\footnotemark[1]
&0.054(5)\footnotemark[1]&0.61\footnotemark[2]\\
Fe$_{3}$Al&&8$c$&0.05\footnotemark[3]&& \\
&&4$b$&0.19\footnotemark[3]&& \\
\hline
Fe$_{2}$VGa&0.577\footnotemark[2]&8$c$&0.161(5)\footnotemark[1]
&0.054(5)\footnotemark[1]&0.59\footnotemark[2]\\
Fe$_{3}$Ga&&8$c$&0.18\footnotemark[4]&& \\
&&4$b$&0.28\footnotemark[4]&& \\
\\
\end{tabular}
\end{ruledtabular}
\footnotetext[1]{This work.}
\footnotetext[2]{Ref.~\onlinecite{lue01}.}
\footnotetext[3]{Ref.~\onlinecite{lin81}.}
\footnotetext[4]{Ref.~\onlinecite{kawamiya72}.}
\end{table}

\section{Discussion}

Previously, the temperature dependences of the $^{27}$Al
and $^{51}$V linewidths in Fe$_{2}$VAl were fit using the analytical
results of Walstedt and Walker\cite{walstedt74}
for dipolar broadening due to random local-moment spins.
Taking the local moments to be antisite defects of 3.7 $\mu_{B}$
as obtained from specific heat measurements,\cite{lue99}
this fit yielded a defect concentration of $c$ = 0.0045,\cite{lue01}
per formula unit, 
which compares favorably with the value $c$ = 0.0037 obtained
from specific heat. In the dilute limit, the lineshape
for this mechanism will be Lorentzian, with identical 
widths for $^{27}$Al and $^{51}$V, as indeed observed 
in Fe$_{2}$VAl.\cite{lue01}  These results gives confidence that 
the broadening mechanism and the defect concentration are
well understood.  Note that several groups have identified
the presence of large-moment clusters in addition to the 
antisite local moments in Fe$_{2}$VAl,\cite{lue01,feng01}
however these turn out to have a much smaller contribution to 
the Fe$_{2}$VAl NMR linewidth.\cite{lue01}

For dipolar broadening in NMR, in the dilute limit one finds 
that the
Lorentzian line has a width proportional to 
$c$.\cite{walstedt74} 
From the data of Fig.~\ref{fig:fig1} we find that the $x$ = 0.10
sample exhibits a half-width 17 times larger than that of
the $x$ = 0 sample.  Nominally this implies an antisite
concentration for $x$ = 0.10 of $c$ = 0.077 per formula unit, 
and although this 
value is outside the dilute limit where the analytic 
expressions are no longer exact, this provides an approximate
estimate of the density of local defects.  This sample
contains $c$ = 0.10 excess Fe per formula unit, so the model 
described here shows that 
a sizable portion (possibly all) of the
excess Fe is distributed as random defects on V sites, 
rather than clustered within the sample.

NMR spin-lattice relaxation due to an uncorrelated population 
of local moments has been treated by several 
authors.\cite{lowe68,tse68,narath72,sung73,furman95} 
If nuclear spin-diffusion is not important, the relaxation
function is a stretched exponential, 
$s(t) \propto exp-(t/\tau_{1})^{1/2}$, due to an inhomogeneous
distribution of local relaxation rates.
For a concentration $n$ per unit volume of effective moments
$p$, the exponential factor is,\cite{tse68}
\begin{equation}
\tau_{1}^{-1}=0.84 \frac{p \mu_{B}\gamma_{n}n}{\sqrt{\omega}}%
\left(\frac{\omega \tau_{c}}{1+\omega^{2}\tau_{c}^{2}}\right)^{1/2}%
\label{eq:one},
\end{equation}
where $\omega$ is the NMR frequency, and a single Debye-type correlation 
time $\tau_{c}$ has been assumed to apply to the local moments. 
At a temperature
where $\tau_{c} = \omega$, a maximum occurs in Eq.~(\ref{eq:one}),
giving $\tau_{1}^{-1}$ = 0.19 s$^{-1}$ for our $x$ = 0 sample,
using the values $p$ = 3.7 and $n = 7.6\times 10^{19}$ cm$^{-3}$ 
(corresponding to $c$ = 0.0037), values taken from the specific
heat measurement\cite{lue01} and described above. 
A distribution of correlation times will give a somewhat lower
peak value.
In our $T_{1}^{-1}$ measurements, for each temperature the
signal amplitude was measured over one or two decades of recovery
time, and a best fit was made to 
the three-exponential $I$ = 5/2 magnetic relaxation function for 
central transition inversion-recovery.\cite{narath67} For
central-transition measurements, 
the stretched-exponential curve for local-moment relaxation
should be convoluted with the same $I$ = 5/2 multi-exponential.
From numerical plots we found that such a 
convolution is nearly indistinguishable from the three-exponential
recovery curve if $\tau_{1}=T_{1}$ over about one decade of 
recovery time.  The peak value observed for $x$ = 0 (Fig.~\ref{fig:fig2}) 
is $T_{1}^{-1}$ = 0.13 s$^{-1}$, in good agreement with the
value 0.19 s$^{-1}$ obtained from the model
described here.  Therefore, just as for the lineshapes, the
density of local moments obtained independently, and attributed
to antisite defects, is in good quantitative agreement with the
observed $^{27}T_{1}^{-1}$ peak value in the $x$ = 0 sample.

Local-moment relaxation of this type scales with $n$
(Eq.~(\ref{eq:one}), exact in the limit of dilute moments).  
From the data in Fig.~\ref{fig:fig2}, the peak value of
$T_{1}^{-1}$ is found to be 17 times larger for the $x$ = 0.10
sample than for the $x$ = 0 sample.  This is
the same ratio as found for the linewidths, thus giving quantitative
consistency with the model, showing that
the excess Fe act as local moments fluctuating independently
in the paramagnetic regime at this concentration. 
The peaks observed in $T_{1}^{-1}$ coincide with maxima
in the ac susceptibility,\cite{luethesis} and therefore
correspond to magnetic transitions in the samples. The 
relaxation peaks can be attributed to
critical slowing down of the local moment dynamics at these 
transition temperatures. Maxima in the
$\mu$SR longitudinal relaxation rate were also observed by 
Maksimov, {\it et al.}\cite{maksimov01} attributed to the
same mechanism.

For the $x$ = $-$0.05 sample the density of local moments is
smaller than for the $x$ = 0.10 sample; this can be seen from
both the NMR linewidth and the peak value of $^{27}T_{1}^{-1}$.
On the other hand, the transition temperature for this sample is
the largest of the three.  Of course, the local moment density
might be expected to be smaller since there is no Fe excess 
in this sample. Thus, the moment density does not necessarily
correlate with T$_{c}$ in this material. Indeed, as has been
pointed out previously,\cite{maksimov01} the local moment density
is below the percolation limit in all of these samples, and thus
it is not sensible that the magnetic transitions are driven 
solely by the interaction of these moments.
 
Another feature identified in Fe$_{2}$VAl is the presence of
a more dilute set of large moments.\cite{lue01,feng01}
These were shown previously to induce superparamagnetic behavior in
Fe$_{2}$VAl,\cite{lue01} although their contribution to the
NMR linewidth is small.  The spin-lattice relaxation time will
also not be affected by these moments, since they will be
saturated in the measuring magnetic field.  
Since no second phase was observed in electron microprobe 
measurements, we propose that the large moments
are ferromagnetic regions in the Heusler lattice having 
enhanced conduction electron density, perhaps associated
with variations in Fe concentration. 
Indeed, Fe$_{2}$VAl is a semimetal, but with
increasing $x$ becomes ferromagnetic, its metallic bands 
becoming occupied.\cite{lue00,kato00} 
The interaction of locally ferromagnetic
regions can lead to superparamagnetic freezing in
Fe$_{2}$VAl, while with increasing $x$ these regions percolate,
leading to ferromagnetic ordering. According to the 
values obtained earlier,\cite{lue01} 
the 17 K $T_{1}^{-1}$ peak for $x$ = 0 must be due to
field polarization of the antisite spins, while the 
higher-temperature peaks for $x \neq 0$ must be due to
antisite defects which are coupled by band electrons.

With sufficient Fe concentration 
Fe$_{2+x}$V$_{1-x}$Al will approach 
Fe$_{3}$Al, in which the 8$c$ sites carry a
moment, in addition to the 4$b$ sites (antisites for
Fe$_{2}$VAl).\cite{stearns68} M\"ossbauer measurements 
give us a measure of the electronic structure changes
involved in this behavior.
We have found (Table~\ref{tab:table1}) that the isomer shift
for Fe$_{2}$VAl is nearly identical to that of the
corresponding 8$c$ site in Fe$_{3}$Al.  A similar result
is seen for Fe$_{2}$VGa and Fe$_{3}$Ga.  A large change
in average isomer shift was found vs. V substitution by
Shobaki {\it et al.},\cite{shobaki00} but this can be
attributed to a change in Fe site occupation.  Since the most
important contribution to the isomer shift is due to the
on-site $s$-electron concentration, clearly very little
electron transfer into $s$-orbits is involved in the 
substitution of Fe for V.  We conclude that at least in
this regard, the Fe
local electronic structure is not substantially changed
by the change of its neighbors from V to Fe.
We picture the addition of Fe leading to the filling of 
$d$-bands in a manner rather like that of a rigid
band model.

The isomer shift for Fe$_{3}$Ga is significantly different
from that of Fe$_{3}$Al.  This seems surprising given the
close similarity of the two materials.  For example,
the vanadium Knight shifts quoted in Table~\ref{tab:table1}
are nearly identical, and these measure the orbital
susceptibility of the $d$-bands in the two materials.
However, large isomer shifts have also been observed in
Fe$_{3}$Ga$_{4}$.\cite{kobeissi99} Apparently the larger
Ga atom leads to more covalent bonding, thus stabilizing
$p$ and $d$ orbitals relative to $s$.

\section{Conclusions}

A measurement of the $^{27}$Al NMR lineshapes and 
spin-lattice relaxation in
Fe$_{2+x}$V$_{1-x}$Al showed that a consistent measure
of the local moment density and dynamics could be
obtained. For $x$ = 0.10, the local moment density
is approximately equal to the excess Fe concentration.
This density is below the percolation limit, although
the material has a ferromagnetic transition.  The 
development of a metallic band can explain this 
transition.  M\"ossbauer measurements show that 
little change in Fe local electronic structure 
accompanies the change from semimetallic Fe$_{2}$VAl
to ferromagnetic Fe$_{3}$Al, with a similar result
found for Fe$_{2}$VGa. 

\begin{acknowledgments}
This work was supported by the Robert A. Welch Foundation, 
Grant No. A-1526.
\end{acknowledgments}

\bibliography{FeVX}

\begin{thebibliography}{40}
\bibitem[{lue()}]{lue98rem}
\bibinfo{note}{In the previous reference, labels on the vertical axis of figure
  3(a) were a factor of 10 too small. This does not affect the conclusions of
  that paper.}
\end{thebibliography}

\end{document}